\documentclass[a4paper,10pt,twoside]{cpc-hepnp}

\usepackage{multicol}
\usepackage{graphicx}
\usepackage{booktabs}
\usepackage{amssymb,bm,mathrsfs,bbm,amscd}
\usepackage[tbtags]{amsmath}
\usepackage{lastpage}

\begin{document}

\fancyhead[co]{\footnotesize Chinese Physics C }

\footnotetext[0]{Received 2 August 2016}

\title{Quenching of fluorescence for linear alkylbenzene\thanks{Supported by the Strategic Priority Research Program of the Chinese Academy of Sciences (Grant No. XDA10010500), the National Natural Science Foundation of China (Grant No. 11205240, 11575193), the Youth Innovation Promotion Association of CAS (No. 2016153) and in part by the CAS Center for Excellence in Particle Physics (CCEPP).}}

\author{%
      Wentai LUO  $^{1;1)}$
\quad Qian LIU  $^{1;1)}$\email{liuqian@ucas.ac.cn}%
\quad Xiang ZHOU $^{2}$
\quad Xuefeng DING  $^{3}$\\
\quad Yangheng ZHENG  $^{1}$
\quad Yayun DING  $^{4}$
\quad Li ZHOU  $^{4}$\\ 
\quad Jun CAO  $^{4}$
\quad Yifang WANG  $^{4}$
}
\maketitle

\address{%
$^1$ University of Chinese Academy of Sciences, Beijing 100049, China\\
$^2$ Hubei Nuclear Solid Physics Key Laboratory, Key Laboratory of Artificial Micro- and Nano-structures of Ministry of Education, and School of Physics and Technology, Wuhan University, Wuhan 430072, China\\
$^3$ Gran Sasso Science Institute (INFN), L'Aquila 67100 Italy \\
$^4$ Institute of High Energy Physics, Chinese Academy of Science, Beijing, 100049, China\\
}

\begin{abstract}
Linear alkylbenzene (LAB) based liquid scintillator is adopted as the central detector for the Jiangmen Underground Neutrino Observatory (JUNO) liquid scintillator detectors. A quenching factor measurement instrument is designed based on the Compton scattering process. Two different quenchers for the liquid scintillator have been investigated and the result shows that the scintillation light of the JUNO liquid scintillator can be quenched to a level. The emission spectrum with the absence of the quencher is also showing a desired behavior. 
\end{abstract}



\begin{multicols}{2}

Due to its ultra transparency~\cite{attenuationLAB}, high light yield~\cite{lightyieldLAB}, long Rayleigh scattering length~\cite{rayleigh1,rayleigh2,rayleigh3} and chemical low-toxic feature, linear alkylbenzene (LAB) is adopted to be the organic solvent of the liquid scintillator (LS) for the Jiangmen Underground Neutrino Observatory (JUNO). Ever since the unexpectedly large $\theta_{13}$ precisely measured by Dayabay~\cite{dayabay}, JUNO is designed to be a multipurpose neutrino experiment for determination of the neutrino mass hierarchy and measurements of oscillation parameters from the medium baseline vacuum oscillations of reactor antineutrinos \cite{juno1, juno2}. 

About 20,000 tons of LAB based LS will be filled in a spherical vessel as the central detector with a diameter of 35.4 m. The antineutrinos are detected via the inverse beta decay reaction $\bar\nu_e+p\rightarrow e^+ + n$.  LAB is excited by positron annihilation and neutron capture, and releases ultraviolet scintillation light from its de-excitation. The scintillation light is red shifted (e.g. to $430$\,nm) by the primary and secondary wavelength shifter~\cite{shifter}: 2,5-diphenyloxazole (PPO). About ten-thousands of photomultiplier tubes (PMTs) are placed surrounding the central detector to capture the scintillator light, and the quantum efficiency is also optimized arround $430$\,nm.

JUNO is a low count rate experiment, the expected number of events is only few tens per day. To reduce the possible background from the radiation of the environment, such as U/Th/K from the glass shield of PMTs and the surroundings, the design is to place a buffer liquid between the PMT and the central detector. One of the options for the buffer liquid is LAB since the density difference between the central detector and the buffer would be small, which introduces a reduction of the buoyancy. However LAB can be excited by the radiation source and generate scintillation light. This motivates the study for quenching those unfavored scintillation lights. 

The quencher must meet the following criteria: It must not absorb the red-shifted scintillation light from the central detector, otherwise we will lose the neutrino detection efficiency and the energy resolution. Meanwhile, the quencher should not bring any new radioactive sources, and it should also be chemically pure and stable, no corrode for any material of the detector, high enough solubility to the buffer liquid and commercially reasonable.

A wide variety of substances act as quenchers. One of the well known quencher is molecular oxygen~\cite{oxygenQuench}. For LAB production for LS, it is necessary to remove dissolved oxygen and protect the LAB by nitrogen during the LAB purification. For the buffer, even though the oxygen is a good quencher, it will bring the aging problem of LAB. Thus it is not a suitable candidate and rejected.

Aromatic and aliphatic amines are also acting as effective quencher for aromatic hydrocarbons. It's reported that the dipyridyl (2'2-DP or LDP) from pyridine derivatives  and dimethylphthalate (DMP) from phthalate derivatives show chemically stable and no corrode to nylon~\cite{MarkCHEN}. These two are chosen as our candidates for the quencher study.

\section{Quenching process} \label{sec:theory}

The fluorescence, first observed by J.F.W. Herschel in 1845 from a quinine solution in sunlight, describes the emission of light from electronically excited states $A^*$. By adding a quencher, this process $A^*+ Q\rightarrow A+ Q$ can happen, where A is the fluorescence and Q is the quencher. This quenching process can be caused by a dynamical collision and described by the Stern-Volmer equation~\cite{svformular}:

\begin{equation}
\label{quencheq}
\frac{F_{0}}{F} = 1+K_{D}[Q].
\end{equation}

Here $F_{0}$ and $F$ are the fluorescent intensities in the absence and presence of quencher, respectively. $K_{D}$ is the quenching constant, and $[Q]$ is the concentration of quencher.

Quenching can also occur as a result of the formation of a non fluorescent ground-state complex between the fluorophore and quencher. This can also be described by a similar equation as the Stern-Volmer, with $K_{D}$ replaced by $K_{S}$. Here $K_{S}$ represents the quenching constant from the static quenching.

In our case the LS can be quenched both by dynamic collision and by complex formation with the same quencher. This can be described by multiplication of the two processes~\cite{quenchbook}:

\begin{equation}
\frac{F_{0}}{F} = (1+K_{D}[Q])(1+K_{S}[Q]).
\end{equation}

The difference to distinguish these two different quenching process is the lifetime. For dynamic quenching process, the $F_0/F=\tau_0/\tau$, while for static quenching, the lifetime is unperturbed hence $\tau_0/\tau=1$.

During our measurement, there are two fluorophores (PPO and bis-MSB). Assuming the quenching process has two different Stern-Volmer coefficients to these two fuorophores: $K_{a}$ and $K_{b}$, we can have:

\begin{equation}
\label{combquench}
\frac{F_{0}}{F} = \frac{1}{ \frac{f_{a}}{1+K_{a}[Q]} + \frac{1-f_{a}}{1+K_b[Q]} }.
\end{equation} 

Here $f_a$ is the fraction of initial fluorescence from component $F_{0a}$ to the total fluorescence $F_{0}=F_{0a}+F_{0b}$.
 
\section{Experimental setup}

This experiment is designed to measure the light yield of a sample excited by a radioactive source~\cite{lightyieldLAB}. The setup is shown in Fig. \ref{fig:Experiment-Sch}. $^{137}$Cs is chosen as the gamma source, collimated and illuminating on a sample held in a cylinder quartz vessel with a dimension of $50$ mm diameter and $120$ mm height. The inner surface of the vessel is covered with an ESR film which has almost $100\%$ reflectivity around $430$ nm. The cover of the vessel is also masked with this kind of film to ensure no leakage of the scintillation light. 

The Compton scattering events $\gamma+e\rightarrow\gamma'+e'$ are selected to perform this measurement, here the primes denote the final states. To measure the intensity of fluorescence, one PMT-I is placed under the bottom of the LS vessel. The  LS scintillation light excited by the deposited energy of the electron $e'$ is collected by PMT-I and the signal is recorded by a CAEN digitizer DT5751. The scattered gamma  $\gamma'$ is detected by a LaBr$_3$ crystal with a dimension of $25$mm diameter and $25$mm height, which is placed in the same scattering plane at a large scattering angle of $\theta_{scat}\approx150^{\circ}$ and the distance from crystal to LS vessel is $100$ mm. PMT-II is equipped with the LaBr$_3$ crystal and this signal is then recorded by a digitizer and treated as a trigger signal to select Compton scattering events. Both of these two PMTs are XP2020, and this experiment was held in a dark room with room temperature controlled at 23\,$^\circ$C.

\begin{center}
\includegraphics[width=.45\textwidth]{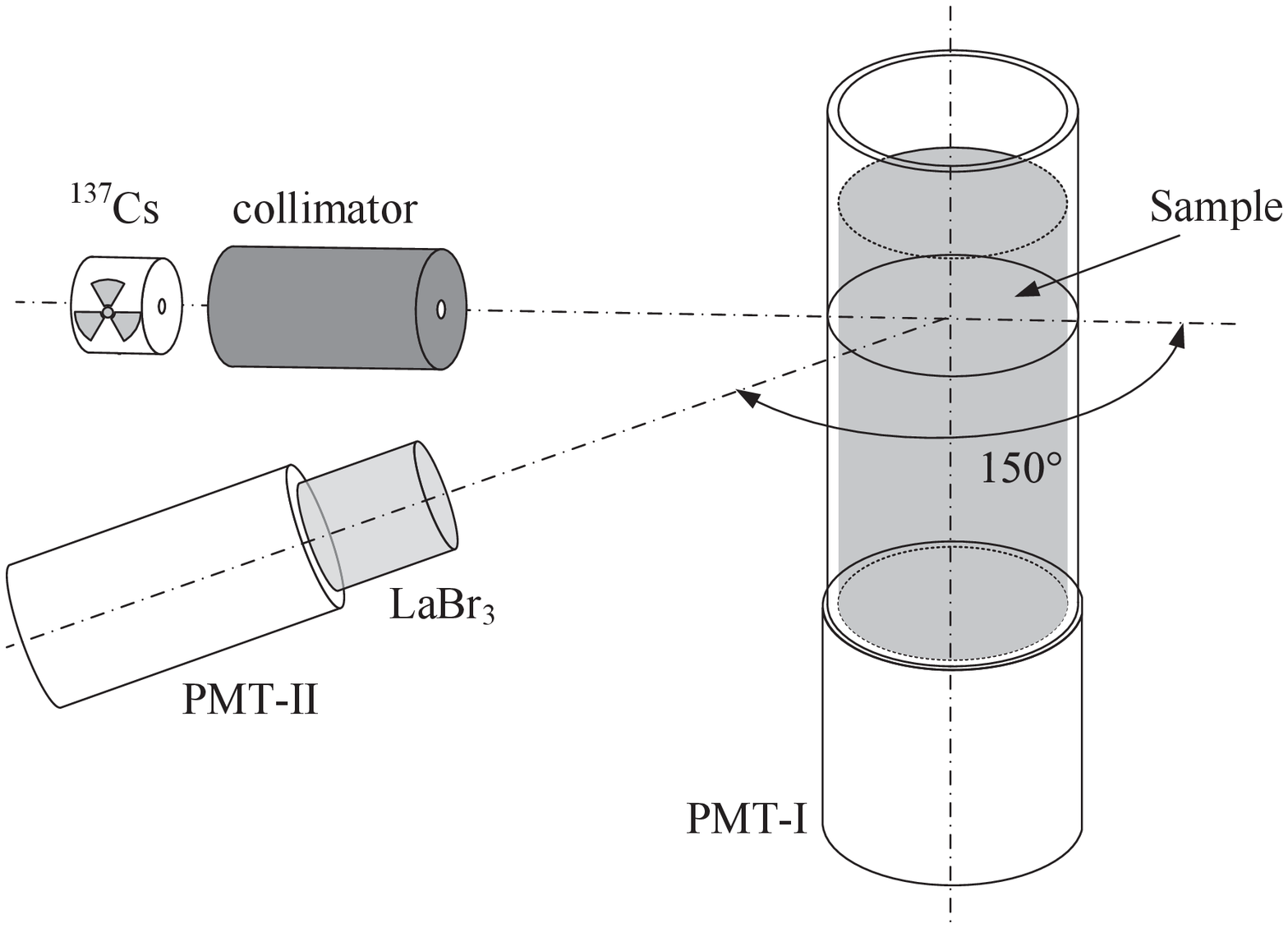}
\figcaption{Sketch of the experimental set-up. \label{fig:Experiment-Sch}}
\end{center}

The scattering angle $\theta_{scat}$ is chosen to be as larger as possible to have almost maximum electron energy deposited in a sample. Since it's an elastic scattering, it is a simple calculation to get the scattered gamma and electron energy at $\theta_{scat}=150^{\circ}$ to be $E_{\gamma'} = 193.7$\,keV and $E_{e'}=468.0$\,keV.

\section{Fluorescence intensity} \label{sec:pc}

A simulation based on Geant4~\cite{Geant4} has been performed. 100 M events are generated and $0.1\%$ events survive by selecting the Compton scattered to the LaBr$_3$ detector. This is consistent with the estimation from the LaBr$_3$ crystal solid angle acceptance. About $45.9\%$ of the selected events are from a single Compton scattering inside the LS.  The energy spectrums deposited in the LaBr$_3$ is shown on the plot (a) in Fig.~\ref{fig:MCsim}. The hollow histogram is the total spectrum and the shaded histogram is the energy from only one Compton scattering inside the LS vessel. It shows that the background events of the spectrums are mainly coming from the multi-Compton scattering. The asymmetry and the spread of the signal is  caused by the acceptance of the detector. By performing a criteria for the deposited energy $188$ keV$<E^{dep}_{\gamma'}<210$ keV (demonstrated as the two red arrows), the deposited energy in the LS for the scattered electron $e'$ is shown on the plot (b) of Fig.\ref{fig:MCsim}. It shows that most of the energy can be deposited in the LS sample and the Compton scattering events can be selected by this criteria.

\begin{center}
\includegraphics[width=.45\textwidth]{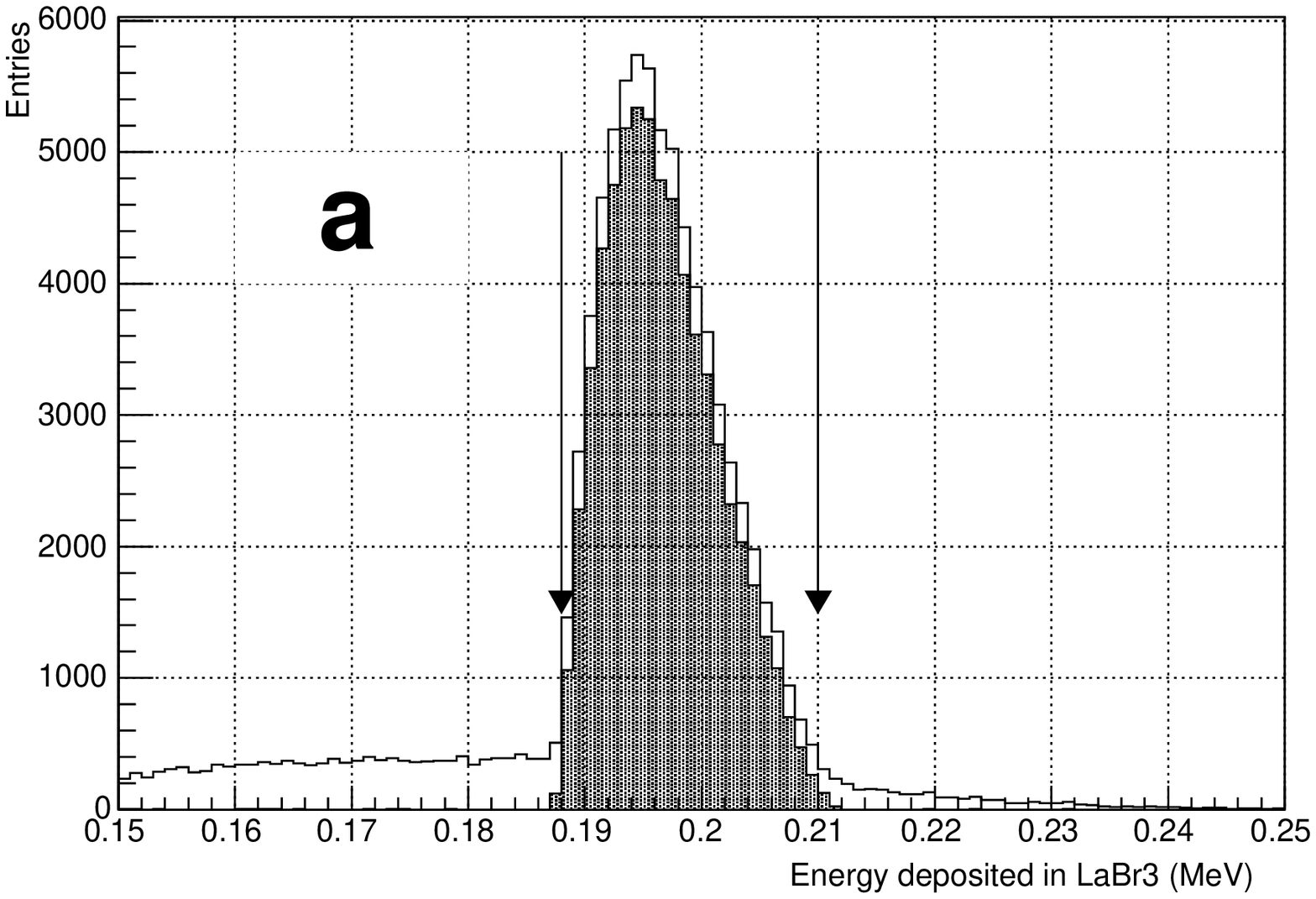}
\includegraphics[width=.45\textwidth]{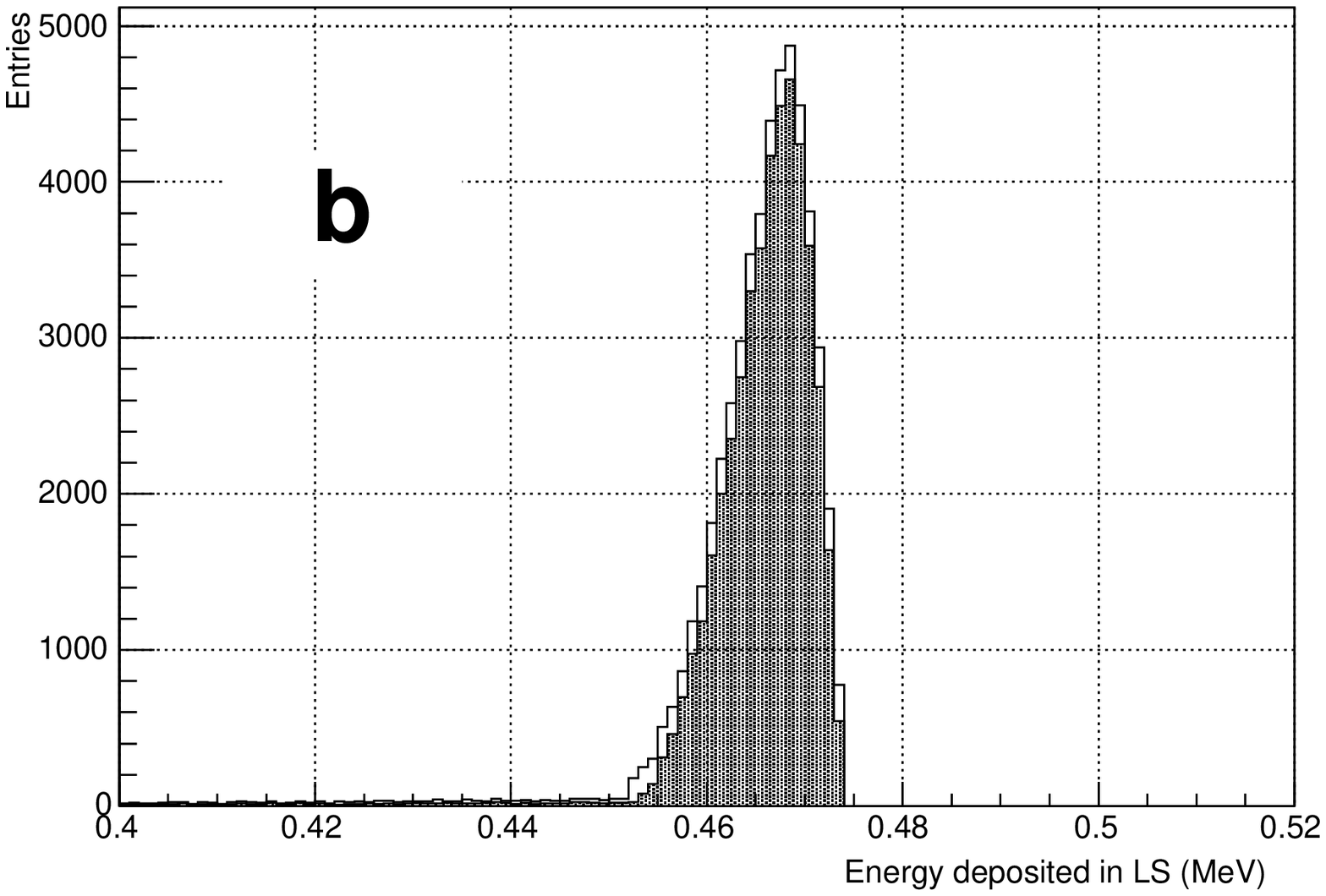}
\figcaption{The plot (a) is the deposited energy of $\gamma'$ spectrum in the LaBr$_3$ detector, and the plot (b) is the deposited energy of $e'$ spectrum in the LS by performing a cut $188$ keV$<E^{dep}_{\gamma'}<210$ keV. The details is described in the text. \label{fig:MCsim}}
\end{center}

The experimental results are given in Fig.~\ref{fig:SPEC}. The plot (a) is from the PMT-II for the scattered gamma $\gamma'$ recorded by the LaBr$_3$ detector. This plot is fitted with a Gaussian plus an Argus to describe the background shape. A criteria of $m-3\sigma<$~E$^{dep}_{\gamma'}<m+3\sigma$ is performed to select the Compton scattering events, and the charge spectrum of the scattered electron $e'$ deposited in the LS is shown on the plot (b). Similar function has been performed to fit the spectrum, and the intensity of fluorescence is described by the mean value of this plot.

\begin{center}
\includegraphics[width=.45\textwidth]{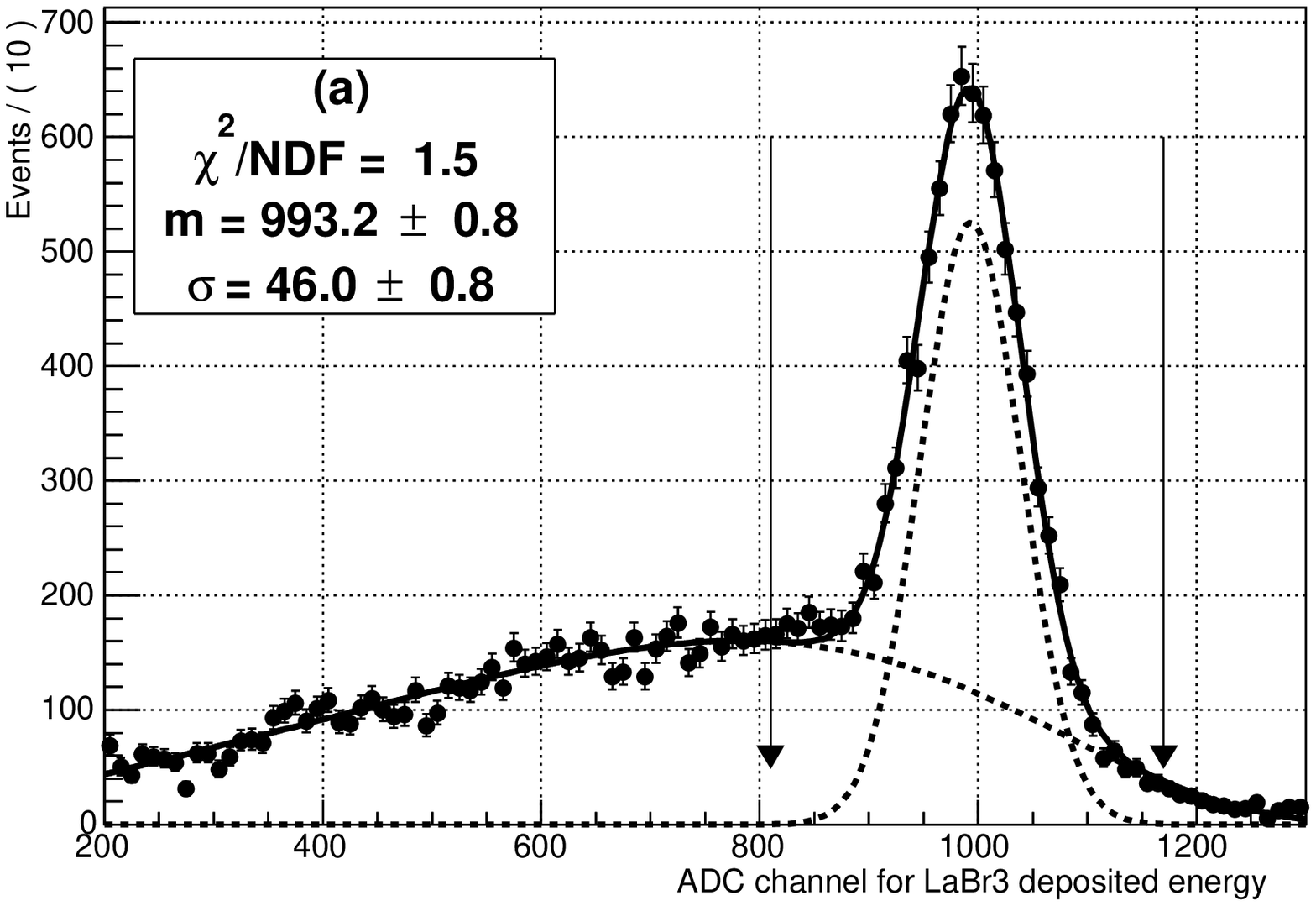}
\includegraphics[width=.45\textwidth]{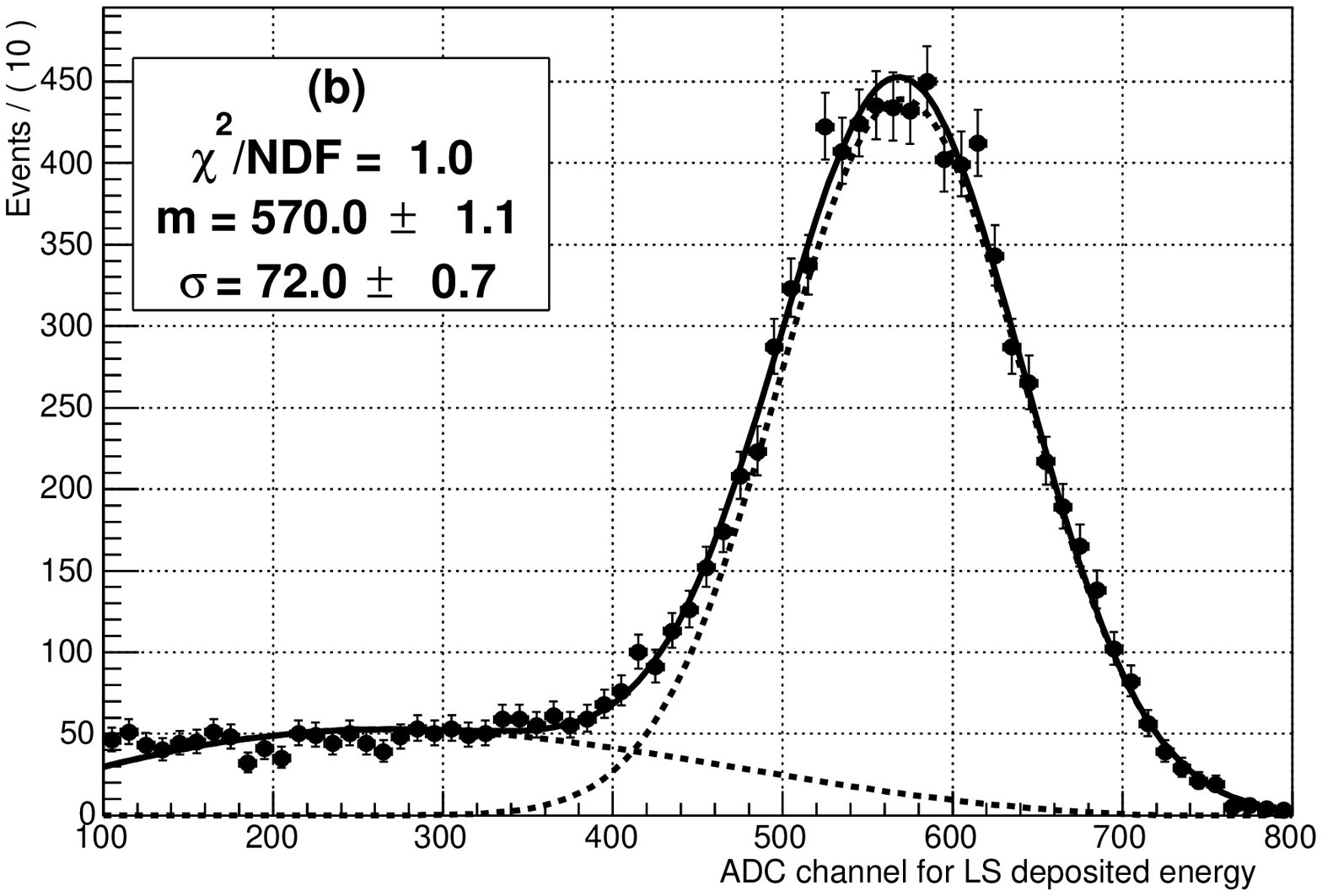}
\figcaption{The plot (a) is the PMT spectrum of the LaBr$_3$ detector, and the plot (b) is the PMT spectrum of the LS. The dots are data, the black line is the fit result, and dashed lines are the contributions of each component. The black arrows demonstrate the cut described in the text. \label{fig:SPEC}}
\end{center}

\section{Result}

The test has been performed with JUNO LS quenched by the LDP and DMP, respectively. The concentration of these two quencher are $2, 5, 7$ g/L ( $0.0103, 0.0257, 0.0360$ Mol/L ). For each test, the LS without quencher is measured first, and then we replace the sample with a new one. A holder is designed for the vessel to guarantee the position of the vessel is the same during this replacement. The position of the LaBr$_3$ detector is fixed during the whole test. The output of this detector is used to check the system stability, which is about $0.5\%$.
 
On the plot (a) of Fig.~\ref{fig:Experiment-Res}, the normalized intensity with different concentration of LDP is shown. Fitted with Eq.~(\ref{quencheq}), it shows that the quenching factor is about $57.4\pm0.9$ Mol$^{-1}$. It has been reported that the pseudocumene quenched by LDP shows the $F_0/F\approx\tau_0/\tau$~\cite{MarkCHEN}. The quenching behavior is mainly a dynamic quenching.

The quenching intensity versus the concentration of DMP is plotted on the plot (b) of Fig.~\ref{fig:Experiment-Res}. It shows that the quenching coefficients of DMP to the two fluorophores are different. By fitting with Eq.~(\ref{combquench}), the quenching coefficients are $70.1$ Mol$^{-1}$ and $2.7$ Mol$^{-1}$, respectively.

\begin{center}
\includegraphics[width=.43\textwidth]{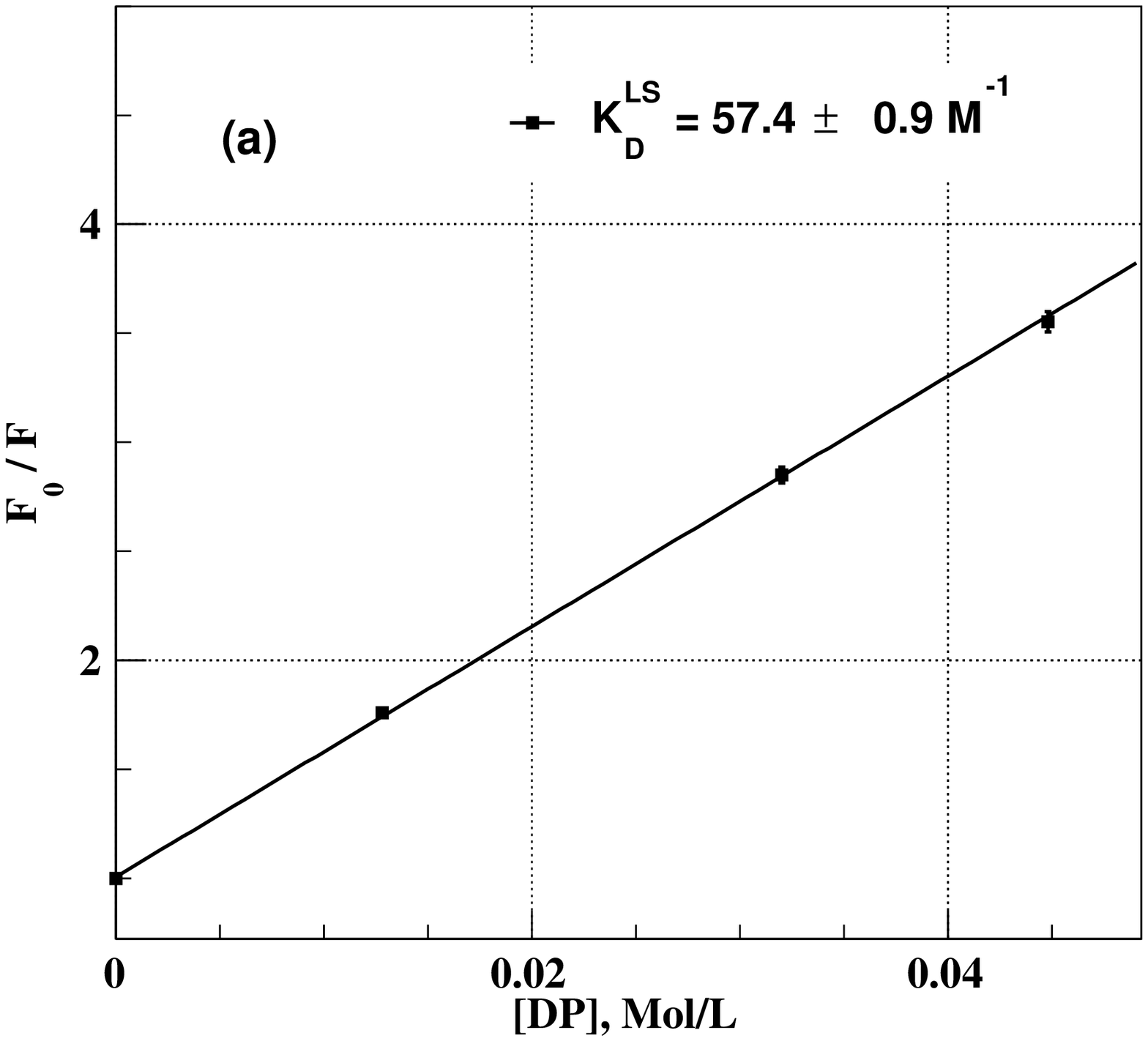}
\includegraphics[width=.43\textwidth]{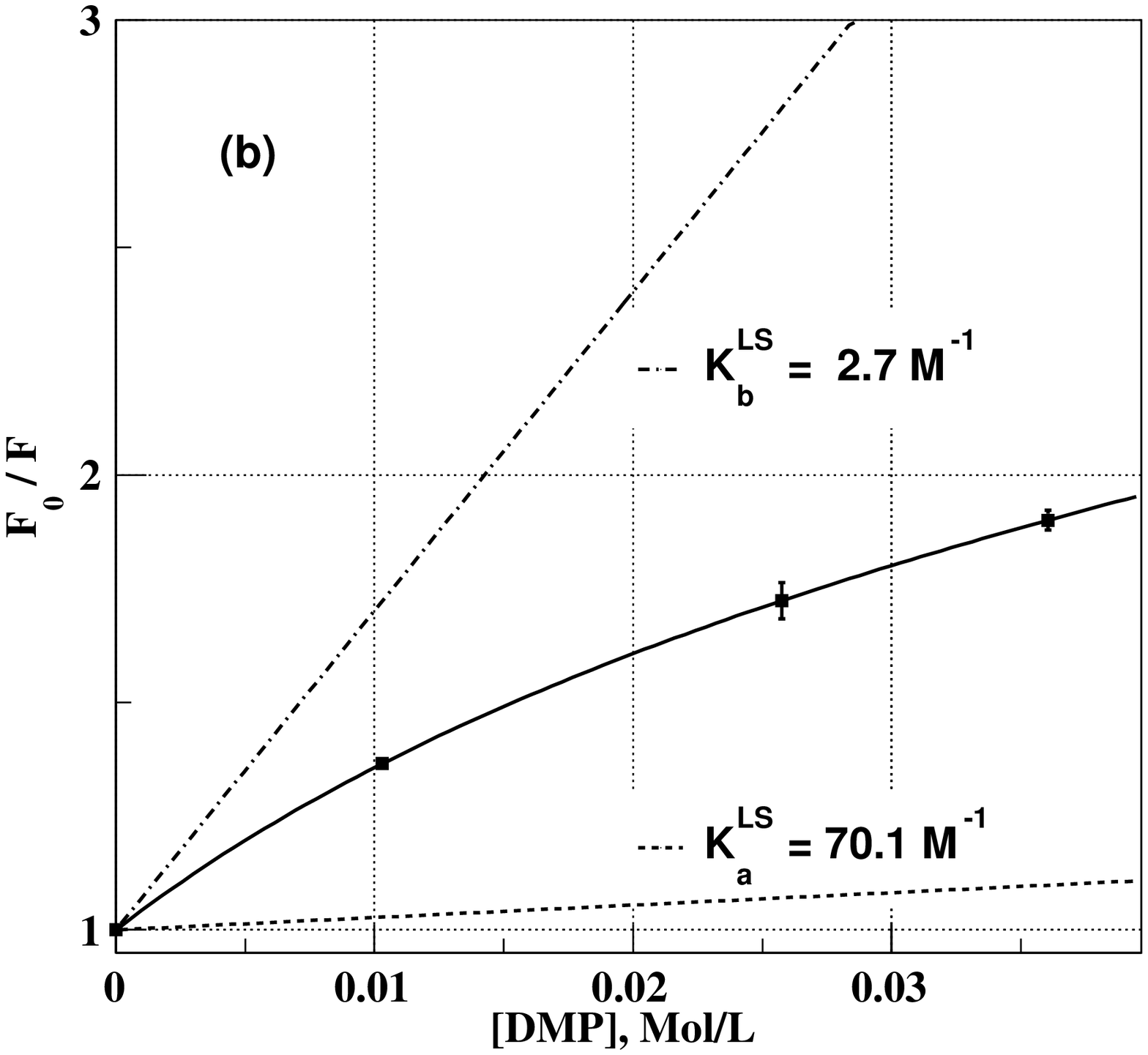}
\figcaption{The plot (a) is the LDP quenching of JUNO LS as observed by fluorescence intensity, the plot (b) is the DMP quenching of LS as observed by fluorescence intensity. The details are described in the text. \label{fig:Experiment-Res}} 
\end{center}

\section{Emission spectrum}

From the previous study, LBP doesn't show any aborption to the light emitted by the pseudocumene~\cite{MarkCHEN}, and the quenching behavior for pseudocumene and JUNO LS is similar. However, DMP shows a different quenching pattern to JUNO LS. The emission spectrum with DMP diluted in the JUNO LS is measured by a fluorescence spectrometer. The samples are excited by $320$ nm and $400$ nm, respectively. The results are shown in Fig.~\ref{fig:emission}.  As discussed above, the quencher should not absorb the emission spectrum of PPO or bis-MSB. In particular the scintillation light around $430$ nm should not be absorbed, otherwise it decreases the anti-neutrino detection efficiency. 

The spectrum has been normalized by the quenching factor from the previous results. It shows that the spectrum shape has a negligible difference by adding the quencher. This indicates that this quencher doesn't absorb the scintillation light in our interesting wavelength range.

\begin{center}
\includegraphics[width=.43\textwidth]{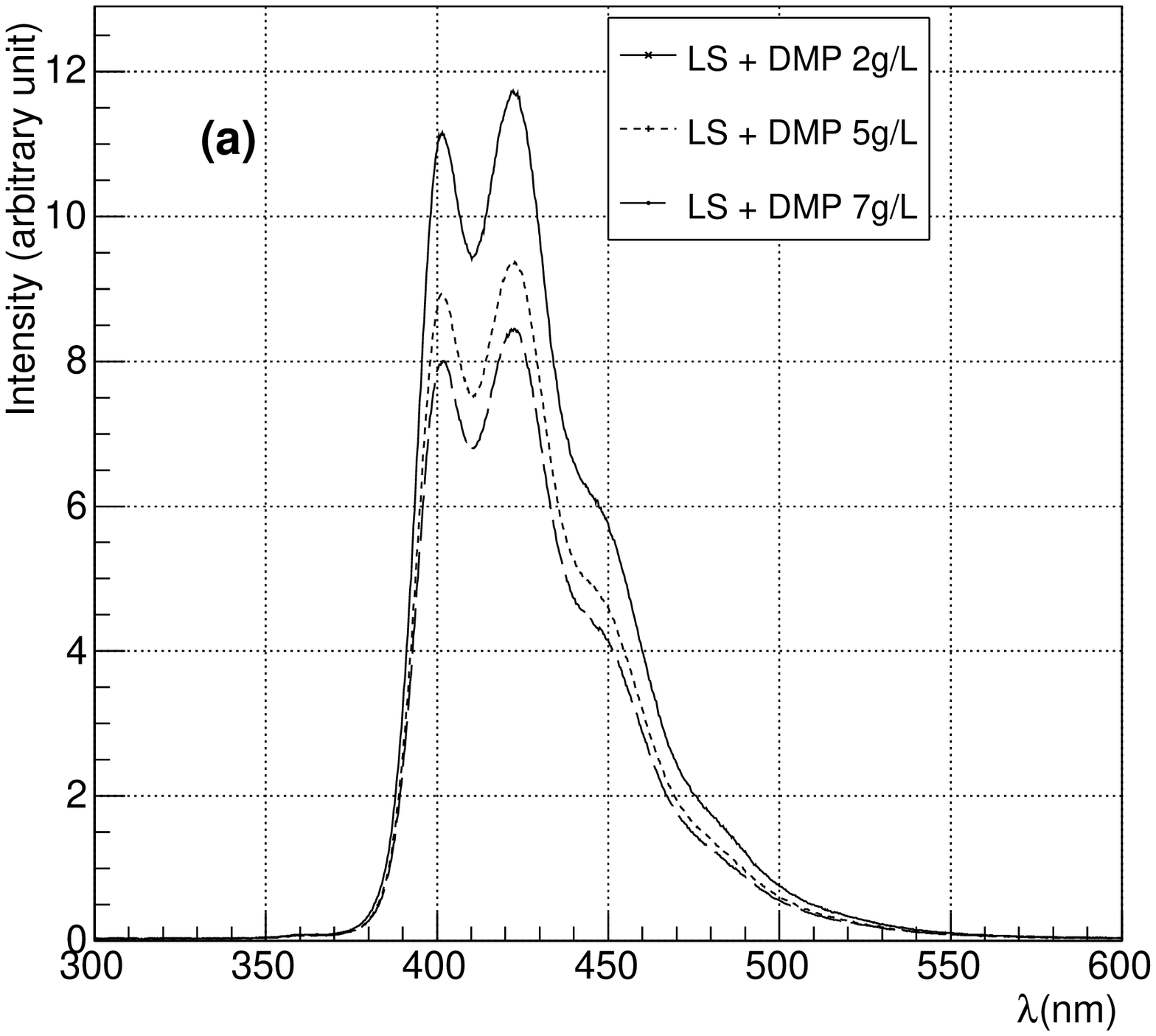}
\includegraphics[width=.43\textwidth]{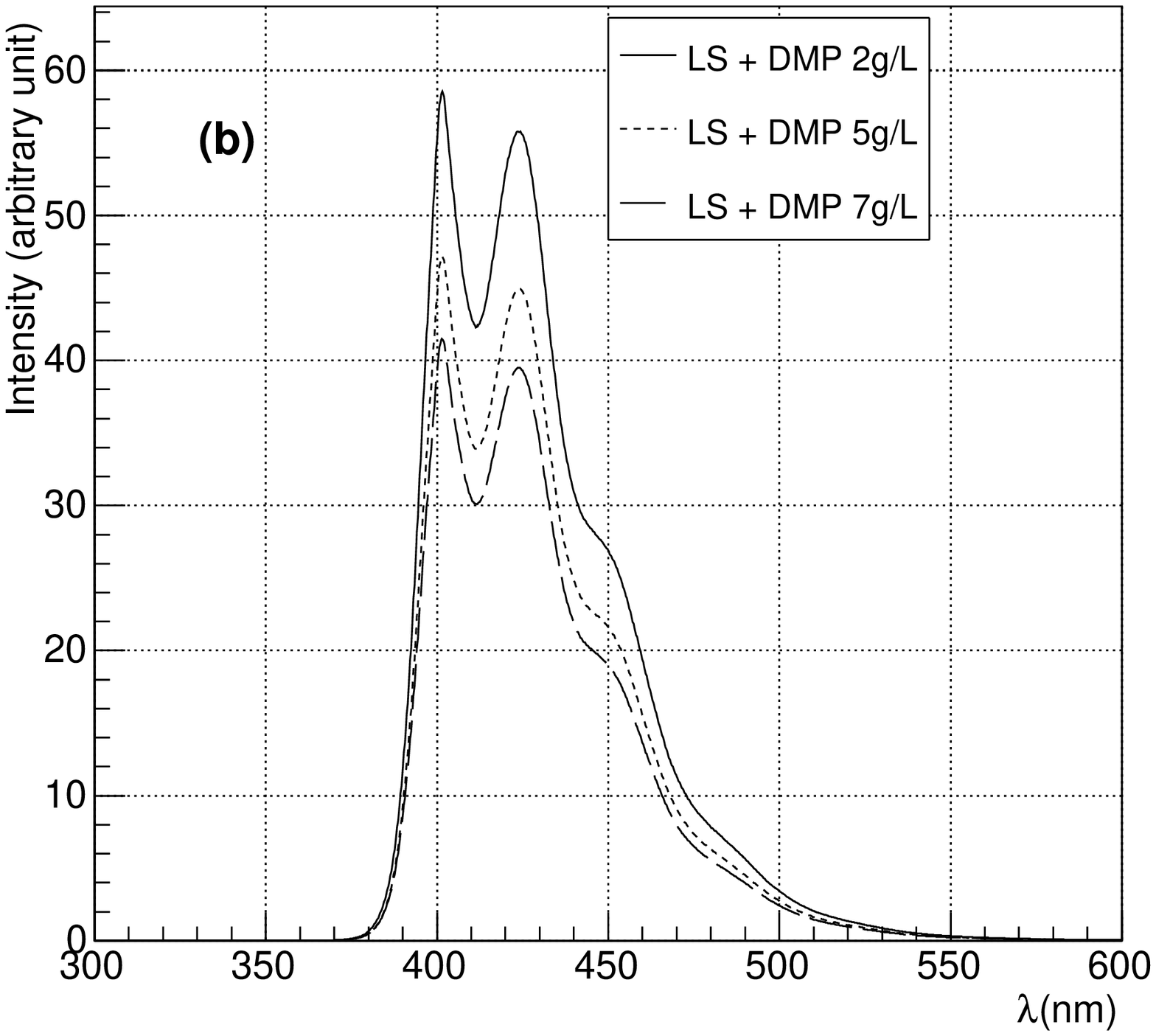}
\figcaption{The plot (a) is the emission spectrum of the JUNO LS with different concentration of DMP quencher excited by $320$ nm. The plot (b) is the same but excited by $400$ nm. \label{fig:emission} }
\end{center}

\section{Discussion and conclusion}

Neutrino experiments are usually low event rate experiments, and it is required to control the background to a limited level. A buffer layer is one of the most effective methods to absorb the radiation of the environment. However the scintillation light from the buffer will be unfavored. Two quenchers have been tested for this purpose. For LDP it is found that the quenching process is mainly a dynamic process and the quenching factor is about $57.4$ Mol$^{-1}$. For DMP it shows a different quenching pattern, the quenching factor is different to the two fluorophores of JUNO LS, one is $2.7$ Mol$^{-1}$ and the other is $70.1$ Mol$^{-1}$. The emission spectrums are also measured for the quencher, and it shows that the quenchers doesn't absorb the scintillation light emission of LS. All of these two quenchers investigated can achieve the desired quenching effect.

\end{multicols}

\vspace{5mm}

\begin{multicols}{2}

\end{multicols}

\clearpage

\end{document}